\documentclass[reprint,superscriptaddress,aps,twocolumn,showpacs]{revtex4-1}
\usepackage{amssymb}
\usepackage{amsmath}
\usepackage{graphicx}
\usepackage{multirow}
\usepackage{hyperref}

\allowdisplaybreaks

\begin{document}
\title{Axial transition form factors of octet baryons in the perturbative chiral quark model}

\author{X. Y. Liu}\email[]{liuxuyang@lnu.edu.cn}
\affiliation{School of Physics, Liaoning University, Shenyang 110036, China}
\affiliation{School of Physics and Center of Excellence in High Energy Physics $\&$ Astrophysics, Suranaree University of Technology, Nakhon Ratchasima 30000, Thailand}
\author{A. Limphirat}\email[]{ayut@g.sut.ac.th}
\author{K. Xu}
\author{Z. Zhao}
\author{K. Khosonthongkee}
\affiliation{School of Physics and Center of Excellence in High Energy Physics $\&$ Astrophysics, Suranaree University of Technology, Nakhon Ratchasima 30000, Thailand}
\author{Y. Yan}
\affiliation{School of Physics and Center of Excellence in High Energy Physics $\&$ Astrophysics, Suranaree University of Technology, Nakhon Ratchasima 30000, Thailand}

\date{\today}

\begin{abstract}
We study the axial transition form factors $G_A^{B\to B'}(Q^2)$ as well as the axial charges $g_A^{B\to B'}$ of the octet baryons in the perturbative chiral quark model~(PCQM) with including both the ground and excited states in the intermediate quark propagators. The PCQM results on the $G_A^{B\to B'}(Q^2)$ and the $g_A^{B\to B'}$ are found in good agreement with the existing experimental data and the lattice-QCD values. The study figures out that the $G_A^{B\to B'}(Q^2)$ for all transitions behave in the dipolelike form, which is dominantly caused by the three-quark core. The meson cloud with the ground-state quark propagator also plays an extremely important role but results in a flat contribution. The excited-state quark propagator contributing to the $G_A^{B\to B'}(Q^2)$ could be regarded as the higher order correction and it is very limited.
\end{abstract}


\maketitle

\section{\label{sec:Intro}Introduction}
The nucleon axial form factor is a very important set of parameters for the investigation of the nucleon internal structure and it is associated with the neutron $\beta$ decay although its measurements are available only for the finite four-momentum transfer $Q^2$ and with large uncertainty~\cite{Bernard:2002}. In the past decade, new values of the axial mass $M_A$ have been extracted from the neutrino-nucleus ($\nu$-$A$) scattering cross section data at several laboratories, such as $M_A=1.05\pm 0.02\pm 0.06$ GeV at NOMAD~\cite{Lyubushkin:2009}, $M_A=1.39\pm 0.11$ GeV at MiniBooNe~\cite{Arevalo:2010}, $M_A=0.99$ GeV at MINER$\nu$A~\cite{Fields:2013,*Fiorentini:2013}, $M_A=1.23^{+0.13}_{-0.09}$ GeV at MINOS~\cite{Adamson:2015} and $M_A=1.26^{+0.21}_{-0.18}$ GeV at T2K~\cite{Abe:2015,*Abe:2016}, and there have been several theoretical works~\cite{Kim:2022,Butkevich:2019,Amaro:2016,Ankowski:2015,Jimenez:2013,Martini:2011,Nieves:2011} to explain the discrepancy between the new data and the traditional nuclear model. Very recently, the nucleon axial form factor has been also estimated in the lattice QCD~\cite{Djukanovic:2022,Alexandrou:2021,Jang:2020,Sufian:2020,Bar:2019,Capitani:2019,Gupta:2017,Alexandrou:2017,Green:2017,Alexandrou:2011,Alexandrou:2007,*Alexandrou:2009}. These lattice QCD simulations are the powerful supplements to the existing experimental data.

Except for the nucleon, however, the direct experimental data on the axial form factors for other baryons are rather limited due to their short lifetimes. The axial charge, which is defined as a value of the axial form factor at zero recoil, has been measured experimentally~\cite{PDG:2020} and also reported in the lattice QCD~\cite{Takahashi:2008,Lin:2008,Lin:2009,Sasaki:2009,Erkol:2010,Alexandrou:2010}. These ``measurements" may also provide the first evidence for the understanding of the baryon electroweak properties and the spontaneous breaking of chiral symmetry.

Pending further experiments, it is of interest to estimate the axial form factors theoretically. In Ref.~\cite{Ramalho:2016}, the $Q^2$ dependence of the axial form factors and the induced pseudoscalar form factors associated with the octet baryon transitions are studied and predicted in the covariant spectator quark model. Moreover, the predictions for the axial transition form factors of the baryon octet to the baryon decuplet are given within the the light-cone QCD sum rules~\cite{Kucukarslan:2016} and the chiral quark-soliton model~\cite{Suh:2022}. In Ref.~\cite{Diaz:2004}, the lowest orbitally excited states of the nucleon and the $\Delta(1232)$ along with the corresponding axial transition form factors are calculated in Poincar\'e covariant constituent quark model. In addition, the baryon axial charge is also evaluated in chiral perturbation theory (ChPT)~\cite{Mendieta:2006,Procura:2007,Jiang:2008,Jiang:2009,Serrano:2014,Ledwig:2014}, QCD sum rule~\cite{Erkol:2011} and quark models~\cite{Yang:2015,Sharma:2009,Faessler:2008,Dahiya:2014,Yang:2015,Choi:2010,Choi2:2010}. These studies inspire us to further predict the axial form factors of the octet baryons by a different method.

The perturbative chiral quark model (PCQM) is an effective tool to study the baryon structures and properties in the low-energy region~\cite{Lyubovitskij1:2001,Lyubovitskij3:2001,Lyubovitskij6:2002,Lyubovitskij2:2002,Lyubovitskij3:2002,Pumsa-ard7:2003,Inoue:2004,Cheedket4:2004,Dong8:2006,Dib9:2006,Faessler5:2008,Liu:2014,Liu:2015,Liu:2018,Liu1:2018,Liu:2019}. In our recent work~\cite{Liu:2018}, we have reextracted the ground and excited-state wave functions for a valence quark in accordance with the Cornell-like potential of the PCQM numerically, and the octet baryon masses have been estimated in the PCQM with considering both the ground and excited states contributing to the quark propagators. Moreover, we have investigated the axial charges of the octet and decuplet baryons~\cite{Liu1:2018} and the charge form factor of the neutron~\cite{Liu:2019}, in which the predetermined ground and excited states were also included into the quark propagators. These PCQM results are found in good agreement with the experimental data and the lattice QCD values, this may indicate that the predetermined quark wave functions are reasonable and credible in the PCQM. The studies also reveal that the meson cloud plays a significant role and the excited states in quark propagators are influential as well. Based on our previous works, we attempt to further study and predict in this work the axial transition form factors for the semileptonic octet baryon decays $B'\to Bl\bar\nu$ in the framework of the PCQM with considering both the ground and excited states in quark propagators. 

The paper is organized as follows. In Sec.~\ref{sec:PCQM&AFF}, we derive and list the PCQM theoretical expressions of the axial transition form factors $G_A^{B\to B'}(Q^2)$, and the numerical results and discussions on the $G_A^{B\to B'}(Q^2)$ are presented in Sec.~\ref{sec:Results}. Finally, we summarize and conclude the work in Sec.~\ref{sec:Summary}.

\section{\label{sec:PCQM&AFF}Axial transition form factors in the PCQM}

In the PCQM, the axial transition form factors $G_A^{B\to B'}(Q^2)$ for the baryon semileptonic decays $B'\to Bl\bar\nu$ are identified  in the Breit frame with the following perturbative expression:
\begin{eqnarray}\label{eq:AFF}
\chi&_{B'_{s}}^\dag&\frac{\vec\sigma}{2}\chi_{B_s}G_A^{B\to B'}(Q^2)\nonumber\\
&=&\,^{B'}\langle \phi_0|\sum_{n=0}^n \frac{\mathrm{i}^n}{n!}\int \delta(t) d^4x d^4x_1\cdots d^4x_n e^{-\mathrm{i}q\cdot x}\nonumber\\
&&\times T[\mathcal{L}_I^W(x_1)\cdots \mathcal{L}_I^W(x_n)\vec A_+(x)]|\phi_0\rangle^B,
\end{eqnarray}
where $\chi_{B_s}$ and $\chi^\dag_{B'_{s}}$ are the baryon spin wave functions in the initial and final states, and $\vec\sigma$ is the baryon spin matrix. The state vector $|\phi_0\rangle^B$ corresponds to the unperturbed three-quark states projected onto the respective baryon states, which are constructed in the framework of the $SU(6)$ spin flavor and $SU(3)$ color symmetry. 

In the right-hand side of Eq.~(\ref{eq:AFF}), the quark-meson interaction Lagrangian $\mathcal{L}_I^W(x)$ is taken in the form 
\begin{eqnarray}\label{eq:WT-int}
\mathcal{L}_I^W(x)&=&\frac{1}{2F}\partial_\mu\Phi_i(x)\bar{\psi}(x)\gamma^\mu\gamma^5
\lambda^i\psi(x)\nonumber\\
&&+\frac{f_{ijk}}{4F^2}\Phi_i(x)\partial_\mu\Phi_j(x)
\bar\psi(x)\gamma^\mu\lambda_k\psi(x),
\end{eqnarray}
where $F=88$ \textrm{MeV} is the pion decay constant, $\Phi_i$ are the octet meson fields, and $\psi(x)$ is the triplet of the $u$, $d$, and $s$ quark fields could be expanded as
\begin{equation}
\psi(x)=\sum_\alpha\left(b_\alpha u_\alpha(\vec{x})\,e^{-iE_\alpha t}+d^\dagger_\alpha\upsilon_\alpha (\vec{x})e^{iE_\alpha t}\right),
\end{equation}
where $b_\alpha$ and $d^\dag_\alpha$ are the single quark
annihilation and antiquark creation operators, and $u_\alpha$ and $\nu_\beta$ are the set of quark and antiquark wave functions in orbits $\alpha$ and $\beta$. Generally, the quark wave functions $u_\alpha(\vec x)$ may be expressed as
\begin{equation}\label{eq:WF}
u_\alpha(\vec{x})=\left(\begin{array}{c}g_\alpha(r)\\\large{i\vec{\sigma}\cdot\hat{x}f_\alpha(r)}
\end{array}\right)\chi_s\chi_f\chi_c.
\end{equation}
Here, $g_\alpha(r)$ and $f_\alpha(r)$ are respectively the upper and lower components of the radial wave functions for a single valence quark. 

The term $A^\mu_+$ in Eq.~(\ref{eq:AFF}) refers to the axial-vector current. For the transition with the $d\leftrightarrow u$, we label it as the variation of the isospin $\Delta I=1$, 
\begin{eqnarray}\label{eq:a-current1}
A_+^\mu |^{\Delta I=1}&=&
\bar\psi\gamma^\mu\gamma^5\frac{\lambda_1+i \lambda_2}{4}\psi
-\frac{f_{1ij}+i f_{2ij}}{4F}\bar\psi\gamma^\mu\lambda_i\psi\Phi_j\nonumber\\
&&+\bar\psi(\hat
Z-1)\gamma^\mu\gamma^5\frac{\lambda_1+i \lambda_2}{4}\psi+o(\Phi_i^2),
\end{eqnarray}
while the transition with the $u\leftrightarrow s$ is labeled as the variation of the strangeness $\Delta S=1$,
\begin{eqnarray}\label{eq:a-current2}
A_+^\mu |^{\Delta S=1}&=&
\bar\psi\gamma^\mu\gamma^5\frac{\lambda_4+i \lambda_5}{4}\psi
-\frac{f_{4ij}+i f_{5ij}}{4F}\bar\psi\gamma^\mu\lambda_i\psi\Phi_j\nonumber\\
&&+\bar\psi(\hat
Z-1)\gamma^\mu\gamma^5\frac{\lambda_4+i \lambda_5}{4}\psi+o(\Phi_i^2),
\end{eqnarray}
The renormalization constant $\hat Z$ in Eqs.~(\ref{eq:a-current1}) and~(\ref{eq:a-current2})  is determined by the nucleon charge conservation condition as
\begin{equation}\label{eq:Z}
\hat Z=1-\frac{1}{3(4\pi F)^2}\sum_{\Phi,\alpha}\int_0^\infty dk \frac{a_\Phi k^4\mathcal{F}_{\Phi NN}^\alpha(k)}{\omega_\Phi(k^2)\big[\omega_\Phi(k^2)+\Delta E_\alpha\big]^2},
\end{equation}
with $\omega_\Phi(k^2)=\sqrt{M_\Phi^2+k^2}$, the constant $a_\pi=9$, $a_K=6$ and $a_\eta=1$. In Eq.~(\ref{eq:Z}), $\Delta E_\alpha=E_\alpha-E_0$ is the energy shift of a single quark, and 
\begin{eqnarray}\hspace{-0.8 cm}
\mathcal{F}_{\Phi NN}^\alpha(k)&=&F_{I\alpha}(k)F_{I\alpha}^\dag(k)-2\omega_\Phi(k^2) F_{I\alpha}(k)F_{II\alpha}^\dag(k)\nonumber\\
&&+\omega_\Phi^2(k^2)F_{II\alpha}(k)F_{II\alpha}^\dag(k),
\end{eqnarray}
\begin{eqnarray}
F_{I\alpha}(k)&=&\int_0^\infty dr r^2\int_\Omega d\Omega\Big[g_0(r)g_\alpha(r)+f_0(r)f_\alpha(r)\cos2\theta\Big]\nonumber\\
&&\times e^{ikr\cos\theta}C_\alpha Y_{l_\alpha0}(\theta,\phi),\label{eq:FI}
\end{eqnarray}
\begin{eqnarray}
\hspace{-10 cm}
F_{II\alpha}(k)&=&\frac{i}{k}\int_0^\infty dr r^2\Big[g_0(r)f_\alpha(r)-g_\alpha(r)f_0(r)\Big]\nonumber\\
&&\times\int_\Omega d\Omega \cos\theta e^{ikr\cos\theta}C_\alpha Y_{l_\alpha0}(\theta,\phi).\label{eq:FII}
\end{eqnarray}
\noindent Here, we define $g_0(r)=g_{1s_{1/2}}(r)C_0 Y_{00}(\theta,\phi)$ and $f_0(r)=f_{1s_{1/2}}(r)C_0 Y_{00}(\theta,\phi)$. The label $\alpha=(nl_\alpha jm)$ in the above equations characterizes the quark state. $C_\alpha$ in Eqs.~(\ref{eq:FI}) and~(\ref{eq:FII}) are the Clebsch-Gordan coefficients $C_\alpha=\langle l_\alpha0\frac{1}{2}\frac{1}{2}|j\frac{1}{2}\rangle$ and $Y_{l_\alpha0}(\theta,\phi)$ is the usual spherical harmonics with $l_\alpha$ being the orbital quantum numbers of the intermediate states $\alpha$. The derivations on the renormalization constant $\hat Z$ could be found in Refs.~\cite{Lyubovitskij1:2001,Lyubovitskij3:2001}.

\begin{figure}[t]
\begin{center}
\includegraphics[width=0.9\columnwidth]{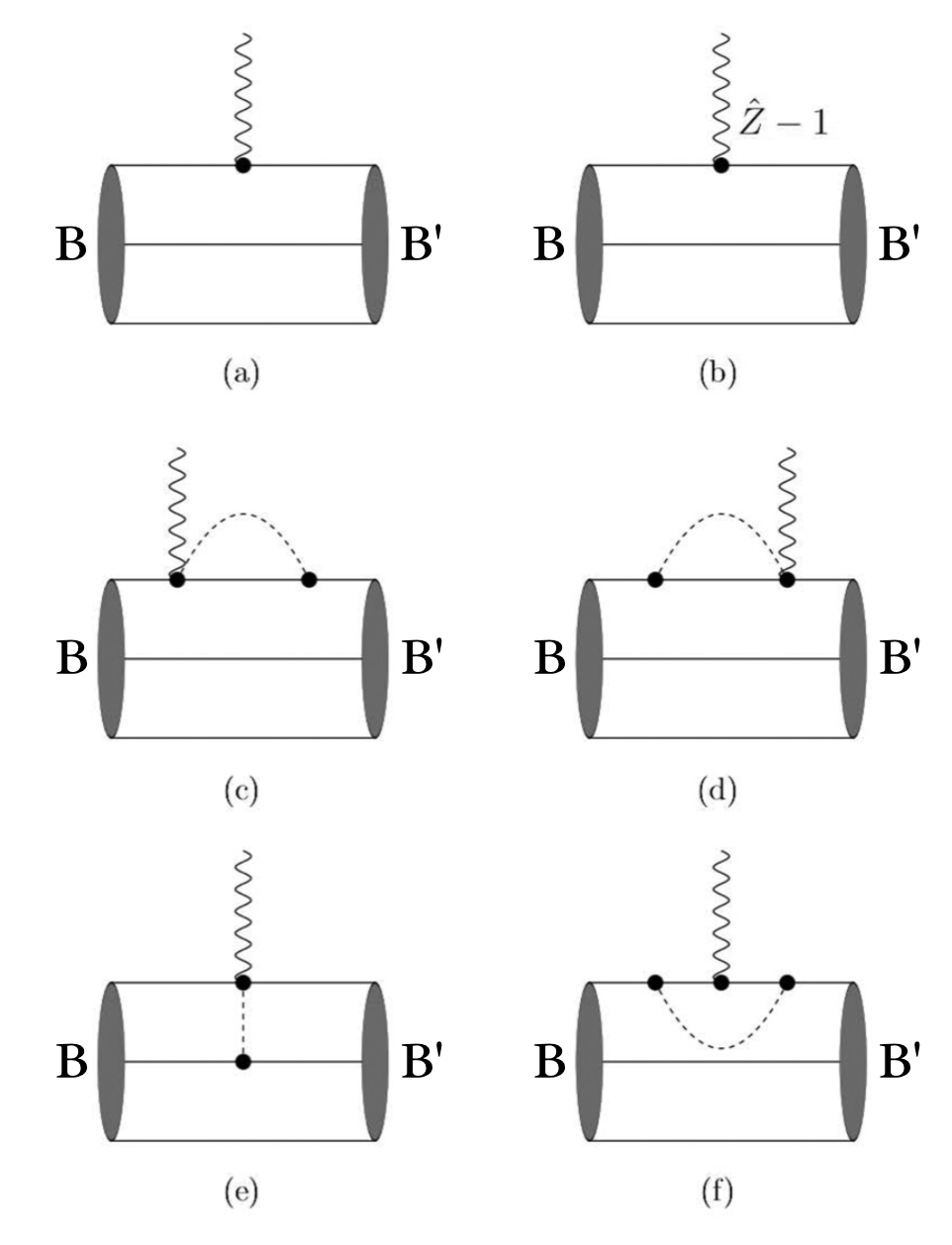}
\caption{\label{fig:AFF}Feynman diagrams contributing to the axial transition form factor of octet baryons : 3q-core leading order (a), 3q-core counterterm (b), self-energy I (c), self-energy II (d), meson exchange (e), and vertex correction (f).}
\end{center}
\end{figure} 

In accordance with the interaction Lagrangian $\mathcal{L}_I^W(x)$ and the axial current $A^\mu_+$, there are six Feynman diagrams contributing to the axial transition form factors as shown in Fig.~\ref{fig:AFF}. The corresponding analytical expressions for the relevant diagrams are derived as follows:\\

\noindent(a) Three-quark core leading-order (LO) diagram
\begin{eqnarray}\label{eq:LO}
G_A^{B\to B'}(Q^2)\big|_{LO}&=&c_1^{BB'} 2\pi\int_0^\infty drr^2 \int_0^\pi d\theta \sin\theta e^{i Q r \cos\theta}\nonumber\\ 
&&\times  [g_0(r)^2+f_0(r)^2\cos(2\theta)].
\end{eqnarray}

\noindent(b) Three-quark core counterterm (CT) diagram
\begin{equation}
G_A^{B\to B'}(Q^2)\big|_{CT}=(\hat Z-1)G_A^{B\to B'}(Q^2)\big|_{LO}.
\end{equation}

\noindent(c) Self-energy I (SE:I) diagram
\begin{eqnarray}\label{eq:SEI}
G_A^{B\to B'}(Q^2)\big |_{SE:I}&=&-\frac{2}{(4\pi F)^2}\int_0^\infty dk \int_{-1}^1 dxk^4\mathcal{F}_{\Phi\alpha}^{BB'}(k)\nonumber\\
&&\times\big[(1-x^2)F_{III\alpha}(k_-)\nonumber\\
&&+(\frac{Q x}{k}+1-2x^2)F_{IV\alpha}(k_-)\big],\label{eq:SEI}
\end{eqnarray}
where $k_-=\sqrt{k^2+Q^2-2k\sqrt{Q^2}x}$, and
\begin{eqnarray}
\mathcal{F}_{\Phi\alpha}^{BB'}(k)&=&\frac{c_2^{BB'}(\omega_\pi(k^2)F_{II\alpha}(k)-F_{I\alpha}(k))}{\omega_\pi(k^2)(\omega_\pi(k^2)+\Delta E_\alpha)}\nonumber\\
&&+\frac{c_3^{BB'}(\omega_K(k^2)F_{II\alpha}(k)-F_{I\alpha}(k))}{\omega_K(k^2)(\omega_K(k^2)+\Delta E_\alpha)}\nonumber\\
&&+\frac{c_4^{BB'}(\omega_\eta(k^2)F_{II\alpha}(k)-F_{I\alpha}(k))}{\omega_\eta(k^2)(\omega_\eta(k^2)+\Delta E_\alpha)},
\end{eqnarray}
\begin{eqnarray}
F_{III\alpha}(k)&=&\frac{i}{k}\int_0^\infty dr r^2g_0(r)f_\alpha(r)\int_\Omega d\Omega \cos\theta\nonumber\\
&&\times e^{ikr\cos\theta}C_\alpha Y_{l_\alpha0}(\theta,\phi),\label{eq:FIII}\\ \nonumber\\
F_{IV\alpha}(k)&=&\frac{i}{k}\int_0^\infty dr r^2[g_\alpha(r)f_0(r)-g_0(r)f_\alpha(r)]\nonumber\\
&&\times\int_\Omega d\Omega \cos\theta e^{ikr\cos\theta}C_\alpha Y_{l_\alpha0}(\theta,\phi).\label{eq:FIV}
\end{eqnarray}

\noindent(d) Self-energy II (SE:II) diagram
\begin{eqnarray}\label{eq:SEII}
G_A^{B\to B'}(Q^2)\big |_{SE:II}&=&G_A^{B\to B'}(Q^2)\big |_{SE:I}, 
\end{eqnarray}
since the SE I and SE II diagrams obey the T symmetry.

\begin{table*}[t!]
\caption{\label{tab:c-constant} The constants $c_i^{BB'}$ for the axial transition form factors $G_A^{B\to B'}(Q^2)$.}
\begin{ruledtabular}
\begin{tabular}{@{}l|llllllllll}
   \multicolumn{1}{l}{ }&
   \multicolumn{1}{l}{Decay}&
   \multicolumn{1}{c}{{}$c_1^{BB'}$}&
   \multicolumn{1}{c}{{}$c_2^{BB'}$}&
   \multicolumn{1}{c}{{}$c_3^{BB'}$}&
   \multicolumn{1}{c}{{}$c_4^{BB'}$}&
   \multicolumn{1}{c}{{}$c_5^{BB'}$}&
   \multicolumn{1}{c}{{}$c_6^{BB'}$}&
   \multicolumn{1}{c}{{}$c_7^{BB'}$}&
   \multicolumn{1}{c}{{}$c_8^{BB'}$}&
   \multicolumn{1}{c}{{}$c_9^{BB'}$}\\[2pt]
\hline\\[-7pt]
$\Delta I=1$ & $n\to p$ & \large$\phantom{-}\frac{5}{3}$ & \large$\phantom{-}\frac{5}{3}$ & \large$\phantom{-}\frac{5}{6}$ & $\phantom{\large{-}}$ $0$ & $\phantom{\large{-}}$ $8$ & $\phantom{\large{-}}$ $0$ & $\phantom{\large{-}}$ $0$ & \large$\phantom{-}\frac{5}{3}$ & \large$-\frac{5}{9}$\\[7pt]
  
   & $\Sigma^- \to \Sigma^0$ & \large$\phantom{-}\frac{\sqrt{8}}{3}$ & \large$\phantom{-}\frac{\sqrt{8}}{3}$ & \large$\phantom{-}\frac{\sqrt{2}}{3}$ & $\phantom{\large{-}}$ $0$ & $\phantom{\large{-}}$ $0$ & $\phantom{\large{-}}$ $2\sqrt{2}$ & $\phantom{\large{-}}$ $0$ & \large$\phantom{-}\frac{\sqrt{8}}{3}$ & \large$-\frac{\sqrt{8}}{9}$\\[7pt]
   
   & $\Xi^- \to \Xi^0$ & \large$-\frac{1}{3}$ & \large$-\frac{1}{3}$ & \large$-\frac{1}{6}$ & $\phantom{\large{-}}$ $0$ & $\phantom{\large{-}}$ $0$ & \large$-$\small$4$ & $\phantom{\large{-}}$ $0$ & \large$-\frac{1}{3}$ & \large$\phantom{-}\frac{1}{9}$\\[7pt]
   
   & $\Sigma^+ \to \Lambda$ & \large$\phantom{-}\frac{\sqrt{6}}{3}$ & \large$\phantom{-}\frac{\sqrt{6}}{3}$ & \large$\phantom{-}\frac{1}{\sqrt{6}}$ & $\phantom{\large{-}}$ $0$ & \large$\phantom{-}\frac{4\sqrt{6}}{3}$ & \large$\phantom{-}\frac{2\sqrt{6}}{3}$ & $\phantom{\large{-}}$ $0$ & \large$\phantom{-}\frac{\sqrt{6}}{3}$ & \large$-\frac{\sqrt{6}}{9}$\\[7pt]
   
\hline\\[-7pt]
 $\Delta S=1$ & $\Sigma^-\to n$ & \large$\phantom{-}\frac{\sqrt{2}}{3}$ & $\phantom{\large{-}}$ $0$ & \large$\phantom{-}\frac{\sqrt{2}}{4}$ & \large$\phantom{-}\frac{\sqrt{2}}{4}$ & $\phantom{\large{-}}$ $0$ & $\phantom{\large-}$ $4\sqrt{2}$ & $\phantom{\large{-}}$ $0$ & $\phantom{\large{-}}$ $0$ & \large$\phantom{-}\frac{2\sqrt{2}}{9}$\\[7pt]
 
   & $\Sigma^0 \to p$ & \large$\phantom{-}\frac{1}{3\sqrt{2}}$ & $\phantom{\large{-}}$ $0$ & \large$\phantom{-}\frac{1}{4\sqrt{2}}$ & \large$\phantom{-}\frac{1}{4\sqrt{2}}$ & $\phantom{\large{-}}$ $0$ & $\phantom{\large{-}}$ $2\sqrt{2}$ & $\phantom{\large{-}}$ $0$ & $\phantom{\large{-}}$ $0$ & \large$\phantom{-}\frac{\sqrt{2}}{9}$\\[8pt]
   
   & $\Xi^- \to \Sigma^0$ & \large$\phantom{-}\frac{5}{3\sqrt{2}}$ & $\phantom{\large{-}}$ $0$ & \large$\phantom{-}\frac{5}{4\sqrt{2}}$ & \large$\phantom{-}\frac{5}{4\sqrt{2}}$ & $\phantom{\large{-}}$ $0$ & $\phantom{\large{-}}$ $\sqrt{2}$ & $\phantom{\large{-}}$ $3\sqrt{2}$ & $\phantom{\large{-}}$ $0$ & \large$\phantom{-}\frac{5\sqrt{2}}{9}$\\[7pt]
   
   & $\Xi^0 \to \Sigma^+$ & \large$\phantom{-}\frac{5}{3}$ & $\phantom{\large{-}}$ $0$ & \large$\phantom{-}\frac{5}{4}$ & \large$\phantom{-}\frac{5}{4}$ & $\phantom{\large{-}}$ $0$ & $\phantom{\large{-}}$ $2$ & $\phantom{\large{-}}$ $6$ & $\phantom{\large{-}}$ $0$ & \large$\phantom{-}\frac{10}{9}$\\[7pt]
   
   & $\Xi^- \to \Lambda$ & \large$-\frac{1}{\sqrt{6}}$ & $\phantom{\large{-}}$ $0$ & \large$-\frac{\sqrt 6}{8}$ & \large$-\frac{\sqrt 6}{8}$ & $\phantom{\large{-}}$ $0$ & \large$-$\small$\sqrt{6}$ & $\phantom{\large{-}}$ $\sqrt{6}$ & $\phantom{\large{-}}$ $0$ & \large$-\frac{\sqrt{6}}{9}$\\[7pt]
   
   & $\Lambda\to p$ & \large$\phantom{-}\frac{\sqrt{6}}{2}$ & $\phantom{\large{-}}$ $0$ & \large$\phantom{-}\frac{3\sqrt{6}}{8}$ & \large$\phantom{-}\frac{3\sqrt{6}}{8}$ & $\phantom{\large{-}}$ $0$ & $\phantom{\large{-}}$ $2\sqrt{6}$ & $\phantom{\large{-}}$ $0$ & $\phantom{\large{-}}$ $0$ &  \large$\phantom{-}\frac{\sqrt{6}}{3}$\\[5pt]
\end{tabular}
\end{ruledtabular}
\end{table*}

\begin{widetext}
\noindent(e) Exchange (EX) diagram
\begin{eqnarray}
G_A^{B\to B'}(Q^2)\big|_{EX} =-\frac{1}{(4\pi F)^2}\int_0^\infty dkk^4 \int_{-1}^1 dx (1-x^2)F_{I0}(k)F_{II0}(k_-)\bigg[\frac{c_5^{BB'}}{\omega_\pi^2(k^2)}+\frac{c_6^{BB'}}{\omega_K^2(k^2)}+\frac{c_7^{BB'}}{\omega_\eta^2(k^2)}\bigg].
\end{eqnarray}

\noindent(f) Vertex-correction (VC) diagram
\begin{eqnarray}\label{eq:VC}
G_A^{B\to B'}(Q^2)\big|_{VC}=\frac{1}{3(4\pi F)^2}\int_0^\infty dk k^4 \mathcal{F}_{\Phi\alpha\beta}^{BB'}(Q^2)[\mathcal{A}_{\alpha\beta}(Q^2)+2\mathcal{B}_{\alpha\beta}(Q^2)+\mathcal{C}_{\alpha\beta}(Q^2)],
\end{eqnarray}
with
\begin{eqnarray}
\mathcal{F}_{\Phi\alpha\beta}^{BB'}(k)&=&c_8^{BB'}\cdot\frac{F_{I\alpha}(k)F_{I\beta}^\dag(k)
-\omega_\pi(k^2) \big[F_{I\alpha}(k)F_{II\beta}^\dag(k)+F_{II\alpha}(k)F_{I\beta}^\dag(k)\big]
+\omega_\pi^2(k^2)F_{II\alpha}(k)F_{II\beta}^\dag(k)}{\omega_\pi(k^2)
[\omega_\pi(k^2)+\Delta\mathcal{E}_\alpha][\omega_\pi(k^2)+\Delta\mathcal{E}_\beta]}\nonumber\\
&&+c_9^{BB'}\cdot\frac{F_{I\alpha}(k)F_{I\beta}^\dag(k)
-\omega_\eta(k^2) \big[F_{I\alpha}(k)F_{II\beta}^\dag(k)+F_{II\alpha}(k)F_{I\beta}^\dag(k)\big]
+\omega_\eta^2(k^2)F_{II\alpha}(k)F_{II\beta}^\dag(k)}{\omega_\eta(k^2)
[\omega_\eta(k^2)+\Delta\mathcal{E}_\alpha][\omega_\eta(k^2)+\Delta\mathcal{E}_\beta]},\\
\mathcal{A}_{\alpha\beta}(Q^2)&=&\int_0^\infty dr r^2[g_\alpha(r)g_\beta(r)-f_\alpha(r)f_\beta(r)]\int_\Omega d \Omega e^{iQr\cos\theta}\big[C_\alpha C_\beta Y_{l_\alpha0}(\theta,\phi)Y_{l_\beta0}(\theta,\phi)-D_\alpha D_\beta Y_{l_\alpha1}^\ast(\theta,\phi)Y_{l_\beta1}(\theta,\phi)\big],\nonumber\\ \\
\mathcal{B}_{\alpha\beta}(Q^2)&=&\int_0^\infty dr r^2f_\alpha(r)f_\beta(r)\int_\Omega d \Omega e^{iQr\cos\theta} \cos^2\theta\big[C_\alpha C_\beta Y_{l_\alpha0}(\theta,\phi)Y_{l_\beta0}(\theta,\phi)-D_\alpha D_\beta Y_{l_\alpha1}^\ast(\theta,\phi)Y_{l_\beta1}(\theta,\phi)\big],\\
\mathcal{C}_{\alpha\beta}(Q^2)&=&\int_0^\infty dr r^2f_\alpha(r)f_\beta(r)\int_\Omega d \Omega e^{iQr\cos\theta}\sin 2\theta\big[C_\alpha D_\beta Y_{l_\alpha0}(\theta,\phi)Y_{l_\beta1}(\theta,\phi)e^{-i\phi}+D_\alpha C_\beta Y_{l_\alpha1}^\ast(\theta,\phi)Y_{l_\beta0}(\theta,\phi)e^{i\phi}\big],\nonumber\\
\end{eqnarray}
\end{widetext}
where $D_\alpha$ are also the Clebsch-Gordan coefficients, and $D_\alpha=\langle l_\alpha1\frac{1}{2}-\frac{1}{2}|j\frac{1}{2}\rangle.$\\

In Table~\ref{tab:c-constant}, we list the constants $c_i^{BB'}$ which are the matrix elements of the flavor and spin operators between baryon states and determined by the projection technique. For the one-body projection
\begin{equation}
\chi^\dagger_{f'}\chi^\dagger_{s'}I^{f'f}J^{s's}\chi^\dagger_{s}\chi^\dagger_{f}\xrightarrow{Proj.}\langle B'|\sum_{i=1}^3(IJ)^{(i)}|B\rangle,\label{eq:onebody}
\end{equation}
Analogously, for the two-body projection, we have
\begin{eqnarray}
&&\chi^\dagger_{f'}\chi^\dagger_{s'}I^{f'f}_1J^{s's}_1\chi^\dagger_{s}\chi^\dagger_{f}\otimes\chi^\dagger_{k'}\chi^\dagger_{\sigma'}I^{k'k}_2J^{\sigma'\sigma}_2\chi^\dagger_{\sigma}\chi^\dagger_{k}\nonumber\\
&&\xrightarrow{Proj.}\langle B'|\sum_{i\neq j}^3(I_1J_1)^{(i)}\otimes(I_2J_2)^{(j)}|B\rangle.\label{eq:twobody}
\end{eqnarray}

Next, we evaluate the axial transition form factors for the octet baryon semileptonic decays $G_A^{B\to B'}(Q^2)$ numerically with including both the ground and excited states in the quark propagators, and also the calculations are extended to the SU(3) flavor symmetry,  i.e., including $\pi$, kaon and $\eta$-meson cloud contributions. Note that the ground and excited quark wave functions employed in the present work have been extracted by solving the Dirac equation with Cornell-like potential, and have been calibrated and exhibited in our previous works~\cite{Liu:2018,Liu1:2018,Liu:2019}. As we discussed in Ref.~\cite{Liu:2018}, the excited states $1p_{1/2}$, $1p_{3/2}$, $1d_{3/2}$, $1d_{5/2}$, $1f_{5/2}$, $1f_{7/2}$, $2s_{1/2}$, $2p_{1/2}$, $2p_{3/2}$ and $3s_{1/2}$ could be embedded into the intermediate quark propagators. Thus, there is not any free parameter in the following numerical calculations. 

\begin{table*}[t!]
\caption{\label{tab:Axialcharges} Numerical results for the octet baryon transition axial charges $g_A^{B\to B'}$. The experimental data are taken from PDG 2020~\cite{PDG:2020}. For comparison, the predictions from lattice-QCD~\cite{Erkol:2010}, ChTP~\cite{Mendieta:2006}, and various quark models~\cite{Ramalho:2016,Yang:2015,Sharma:2009,Faessler:2008} are listed as well.}
\begin{ruledtabular}
\begin{tabular}{@{}l|lcccccccc}
   \multicolumn{1}{l}{ }&
   \multicolumn{1}{l}{Decay}&
   \multicolumn{1}{c}{{}Expt.~\cite{PDG:2020}}&
   \multicolumn{1}{c}{{}PCQM}&
   \multicolumn{1}{c}{{}Lattice~\cite{Erkol:2010}}&
   \multicolumn{1}{c}{{}Ref.~\cite{Mendieta:2006}}&
   \multicolumn{1}{c}{{}Ref.~\cite{Ramalho:2016}}&
   \multicolumn{1}{c}{{}Ref.~\cite{Yang:2015}}&
   \multicolumn{1}{c}{{}Ref.~\cite{Sharma:2009}}&
   \multicolumn{1}{c}{{}Ref.~\cite{Faessler:2008}}\\[2pt]
\hline\\[-7pt]
  $\Delta I=1$ & $n\to p$ & $1.2756\pm 0.0013$ & $\phantom{-}1.265$ & $\phantom{-}1.314(24)$ & $\phantom{-}1.27$ & $\phantom{-}1.125$ & $\phantom{-}1.269$ & $\phantom{-}1.27$ & $\phantom{-}1.27$\\[7pt]
  
   & $\Sigma^- \to \Sigma^0$ & $\cdots$ & $\phantom{-}0.635$ & $\phantom{-}0.686(15)$ &  \phantom{-} $\cdots$  & $\phantom{-}0.636$ & $\phantom{-}0.439$ & $\phantom{-}0.48$ & \phantom{-} $\cdots$\\[7pt]
   
   & $\Xi^- \to \Xi^0$ & $\cdots$ & $-0.276$ & $-0.300(10)$&  \phantom{-} $\cdots$  & $-0.225$ & $-0.245$ & $-0.31$ & \phantom{-} $\cdots$ \\[7pt]
   
   & $\Sigma^+ \to \Lambda$ & $\cdots$ & $\phantom{-}0.629$ & $\phantom{-}0.655(14)$ & $\phantom{-}0.65$ & $\phantom{-}0.551$ & $\phantom{-}0.794$ & $\phantom{-}0.65$ & $\phantom{-}0.65$\\[7pt]
   
\hline\\[-7pt]
 $\Delta S=1$ & $\Sigma^-\to n$ & $0.340\pm 0.017$ & $\phantom{-}0.356$ & $\phantom{-}0.339(12)$ & $\phantom{-}0.38$ & $\phantom{-}0.225$ & $\phantom{-}0.340$ & $-0.31$ & $\phantom{-}0.26$\\[7pt]
 
   & $\Sigma^0 \to p$ & $\cdots$ & $\phantom{-}0.178$ & \phantom{-} $\cdots$ & \phantom{-} $\cdots$ & \phantom{-} $\cdots$ & \phantom{-} $\cdots$ & \phantom{-} $\cdots$  & \phantom{-} $\cdots$\\[8pt]
   
   & $\Xi^- \to \Sigma^0$ & $\cdots$ & $\phantom{-}0.738$ & $\phantom{-}0.908(19)$ & $\phantom{-}0.87$ & $\phantom{-}0.795$ & $\phantom{-}1.210$ & $\phantom{-}1.27$ & $\phantom{-}0.91$\\[7pt]
   
   & $\Xi^0 \to \Sigma^+$ & $1.22\pm 0.05$ & $\phantom{-}1.044$ & $\phantom{-}1.284(28)$ & $\phantom{-}1.31$ & $\phantom{-}1.125$ & $\phantom{-}1.210$  & $\phantom{-}1.27$ & $\phantom{-}1.28$\\[7pt]
   
   & $\Xi^- \to \Lambda$ & $-0.25\pm 0.05$ & $-0.229$ & $-0.274(08)$ & $\phantom{-}0.14$ & $-0.276$ & $-0.250$  & $-0.21$ & $\phantom{-}0.32$\\[7pt]
   
   & $\Lambda\to p$ & $\phantom{-}0.718\pm 0.015$ & $\phantom{-}0.758$ & $\phantom{-}0.632(14)$ & $\phantom{-}0.90$ & $\phantom{-}0.827$ & $\phantom{-}0.718$  & $\phantom{-}0.74$ & $\phantom{-}0.94$\\[5pt]
\end{tabular}
\end{ruledtabular}
\end{table*}

\section{\label{sec:Results}Numerical results and discussion}

In this section, we first evaluate the octet baryon transition axial charge $g_A^{B\to B'}$,
\begin{equation}
g_A^{B\to B'}=G_A^{B\to B'}(Q^2)|_{Q^2=0},
\end{equation}
and present it in Table~\ref{tab:Axialcharges}. To be compared, we also list the available experimental data~\cite{PDG:2020}, and compile the prediction values from lattice-QCD~\cite{Erkol:2010}, ChPT~\cite{Mendieta:2006} and various quark models~\cite{Ramalho:2016,Yang:2015,Sharma:2009,Faessler:2008} in Table~\ref{tab:Axialcharges}. 

In the case with $\Delta I=1$, it is clear that the PCQM result of the $g_A^{n\to p}$ is in good agreement with the experimental value, and the results of the $g_A^{B\to B'}$ for the $\Sigma^-\to \Sigma^0$, $\Xi\to\Xi^0$ and $\Sigma^+\to\Lambda$ decay processes are slightly smaller than the lattice-QCD estimates. Note that the pion with mass $m_\pi \approx500$ MeV was used in the lattice-QCD simulations in Ref.~\cite{Erkol:2010}, but the PCQM calculations employed the meson with the physical mass. Thus we may regard our results as reasonable. Also, the PCQM yields very similar results of the $g_A^{B\to B'}$ with ChTP~\cite{Mendieta:2006}, covariant spectator quark model~\cite{Ramalho:2016}, chiral soliton model~\cite{Yang:2015},  and chiral constituent quark model~\cite{Sharma:2009} and Lorentz covariant chiral quark model~\cite{Faessler:2008} as shown in Table~\ref{tab:Axialcharges}.

\begin{table}[b!]
\caption{\label{tab:LO&Loop} Numerical results for the $g_A^{B\to B'}$ divided into 3q-core (LO) and meson cloud (loop) contributions.}
\begin{ruledtabular}
\begin{tabular}{@{}l|lccc}
   \multicolumn{1}{l}{}&
   \multicolumn{1}{l}{Decay}&
   \multicolumn{1}{c}{$\phantom{-}$3q-core}&
   \multicolumn{1}{c}{$\phantom{-}$Meson}&
   \multicolumn{1}{c}{$\phantom{-}$Total}\\[3pt]
\hline\\[-10pt]
  $\Delta I=1$ & $n\to p$ & $\phantom{-}0.884$ & $\phantom{-}0.381$ & $\phantom{-}1.265$\\[2pt]
   & $\Sigma^- \to \Sigma^0$ & $\phantom{-}0.500$ & $\phantom{-}0.135$ & $\phantom{-}0.635$ \\[2pt]
   & $\Xi^- \to \Xi^0$  & $-0.177$ & $-0.099$ & $-0.276$\\[2pt]
   & $\Sigma^+ \to \Lambda$ & $\phantom{-}0.433$ & $\phantom{-}0.196$& $\phantom{-}0.629$\\[3pt]
   \hline\\[-10pt]
 $\Delta S=1$ & $\Sigma^-\to n$ & $\phantom{-}0.250$ & $\phantom{-}0.106$ & $\phantom{-}0.356$\\[2pt]
   & $\Sigma^0 \to p$ & $\phantom{-}0.125$ & $\phantom{-}0.053$ & $\phantom{-}0.178$ \\[2pt]
    & $\Xi^- \to \Sigma^0$ & $\phantom{-}0.625$ & $\phantom{-}0.113$ & $\phantom{-}0.738$ \\[2pt]
    & $\Xi^0 \to \Sigma^+$ & $\phantom{-}0.884$ & $\phantom{-}0.160$ & $\phantom{-}1.044$ \\[2pt]
    & $\Xi^- \to \Lambda$ & $-0.217$ & $-0.013$ & $-0.229$ \\[2pt]
    & $\Lambda\to p$ & $\phantom{-}0.650$ & $\phantom{-}0.108$ & $\phantom{-}0.758$ \\[2pt]
\end{tabular}
\end{ruledtabular}
\end{table}

For the case with $\Delta S=1$, one may see that our results are comparable with the experimental data in existence, especially, the ratio $g_A^{\Lambda\to p}/g_A^{\Sigma^-\to n}=2.129$ consists with the experimental value and gives a clear improvement over the results in Refs.~\cite{Sharma:2009} and~\cite{Ledwig:2008}. As shown in Table~\ref{tab:Axialcharges}, we produce the different results for two transitions $\Xi^0\to \Sigma^+$ and $\Xi^-\to \Sigma^0$, and the ratio $g_A^{\Xi^0\to \Sigma^+}/g_A^{\Xi^-\to \Sigma^0}=1.414$ is the same as findings in the lattice-QCD~\cite{Erkol:2010} and other approaches~\cite{Ramalho:2016,Faessler:2008}. Moreover, we also predict the $\Sigma^0\to p$ transition axial charge $g_A^{\Sigma^0\to p}=0.178$, which is much closer to one determined by the cloudy bag model~\cite{Kubodera:1985}.

\begin{table}[b!]
\caption{\label{tab:Meson} Numerical results of the $\pi$, $K$ and $\eta$ mesons contributing to the $g_A^{B\to B'}$ individually.}
\begin{ruledtabular}
\begin{tabular}{@{}l|lcccc}
   \multicolumn{1}{l}{}&
   \multicolumn{1}{l}{Decay}&
   \multicolumn{1}{c}{$\phantom{-}\pi$}&
   \multicolumn{1}{c}{$\phantom{-}K$}&
   \multicolumn{1}{c}{$\phantom{-}\eta$}\\[3pt]
\hline\\[-10pt]
  $\Delta I=1$ & $n\to p$ & $\phantom{-}0.363$ & $\phantom{-}0.027$ & $-0.009$\\[2pt]
   & $\Sigma^- \to \Sigma^0$ & $\phantom{-}0.077$ & $\phantom{-}0.063$ & $-0.005$\\[2pt]
   & $\Xi^- \to \Xi^0$  & $-0.027$ & $-0.074$ & $\phantom{-}0.002$\\[2pt]
   & $\Sigma^+ \to \Lambda$ & $\phantom{-}0.159$ & $\phantom{-}0.041$& $-0.004$\\[3pt]
   \hline\\[-10pt]
 $\Delta S=1$ & $\Sigma^-\to n$ & $-0.034$ & $\phantom{-}0.114$ & $\phantom{-}0.026$\\[2pt]
   & $\Sigma^0 \to p$ & $-0.017$ & $\phantom{-}0.057$ & $\phantom{-}0.013$ \\[2pt]
    & $\Xi^- \to \Sigma^0$ & $-0.085$ & $\phantom{-}0.068$ & $\phantom{-}0.130$ \\[2pt]
    & $\Xi^0 \to \Sigma^+$ & $-0.120$ & $\phantom{-}0.096$ & $\phantom{-}0.184$ \\[2pt]
    & $\Xi^- \to \Lambda$ & $\phantom{-}0.029$ & $-0.057$ & $\phantom{-}0.015$ \\[2pt]
    & $\Lambda\to p$ & $-0.088$ & $\phantom{-}0.129$ & $\phantom{-}0.067$ \\[2pt]
\end{tabular}
\end{ruledtabular}
\end{table}

In Table~\ref{tab:LO&Loop}, we divide the results of the $g_A^{B\to B'}$ into two: 3q-core (LO) and meson cloud (loops) contributions, respectively. It is easy to find that the 3q-core in the PCQM dominates the $g_A^{B\to B'}$, and the meson cloud also retains an enormous influence, contributing around 20\%-30\% to the total values for the most decay processes, but less than 10\% for the $\Xi^- \to \Lambda$ and $\Lambda\to p$ transitions.  

\begin{table}[b!]
\caption{\label{tab:GS&EX} Meson cloud contributions to the $g_A^{B\to B'}$ separated into the ground state and the excited states in the quark propagators.}
\begin{ruledtabular}
\begin{tabular}{@{}l|lcccc}
   \multicolumn{1}{l}{}&
   \multicolumn{1}{l}{Decay}&
   \multicolumn{1}{c}{Ground}&
   \multicolumn{1}{c}{Excited}\\[3pt]
\hline\\[-10pt]
  $\Delta I=1$ & $n\to p$ & $\phantom{-}0.417$ & $-0.036$\\[2pt]
   & $\Sigma^- \to \Sigma^0$ & $\phantom{-}0.156$ & $-0.021$\\[2pt]
   & $\Xi^- \to \Xi^0$  & $-0.106$ & $\phantom{-}0.007$\\[2pt]
   & $\Sigma^+ \to \Lambda$ & $\phantom{-}0.214$ & $-0.018$\\[3pt]
   \hline\\[-10pt]
 $\Delta S=1$ & $\Sigma^-\to n$ & $\phantom{-}0.125$ & $-0.019$\\[2pt]
   & $\Sigma^0 \to p$ & $\phantom{-}0.063$ & $-0.010$\\[2pt]
    & $\Xi^- \to \Sigma^0$ & $\phantom{-}0.161$ & $-0.048$\\[2pt]
    & $\Xi^0 \to \Sigma^+$ & $\phantom{-}0.228$ & $-0.068$\\[2pt]
    & $\Xi^- \to \Lambda$ & $-0.029$ & $\phantom{-}0.016$\\[2pt]
    & $\Lambda\to p$ & $\phantom{-}0.158$ & $-0.050$\\[2pt]
\end{tabular}
\end{ruledtabular}
\end{table}

In order to further illustrate the effect of the meson cloud to the $g_A^{B\to B'}$, we subdivide the meson cloud contributions into the blocks of the $\pi$, $K$ and $\eta$ meson, and list in Table~\ref{tab:Meson}. The numerical results with $\Delta I=1$ reveal that the meson contributions to the $g_A^{B\to B'}$ are caused mainly by the $\pi$ meson and the $K$ meson, while the $\eta$ meson contributes negatively and it could be negligible. Differently, the $K$ and $\eta$ meson contributions to the $g_A^{B\to B'}$ associated with the $\Delta S=1$ transition dominate over the one from the $\pi$ meson, and the $\pi$ meson reduces the $g_A^{B\to B'}$ considerably. Next, we give respectively in Table~\ref{tab:GS&EX} the ground state and the excited states in the quark propagators contributing to the loop Feynman diagrams [Figs.~\ref{fig:AFF}(b)-\ref{fig:AFF}(f)]. Based on the results shown in Table~\ref{tab:GS&EX}, we may point out the fact that the meson cloud contributions to the $g_A^{B\to B'}$ stem mainly from the loop diagrams with the ground state in the quark propagator while the excited-state quark propagators more or less reduce the effect of the meson cloud.

\begin{figure}[t]
\begin{center}
\includegraphics[width=0.48\textwidth]{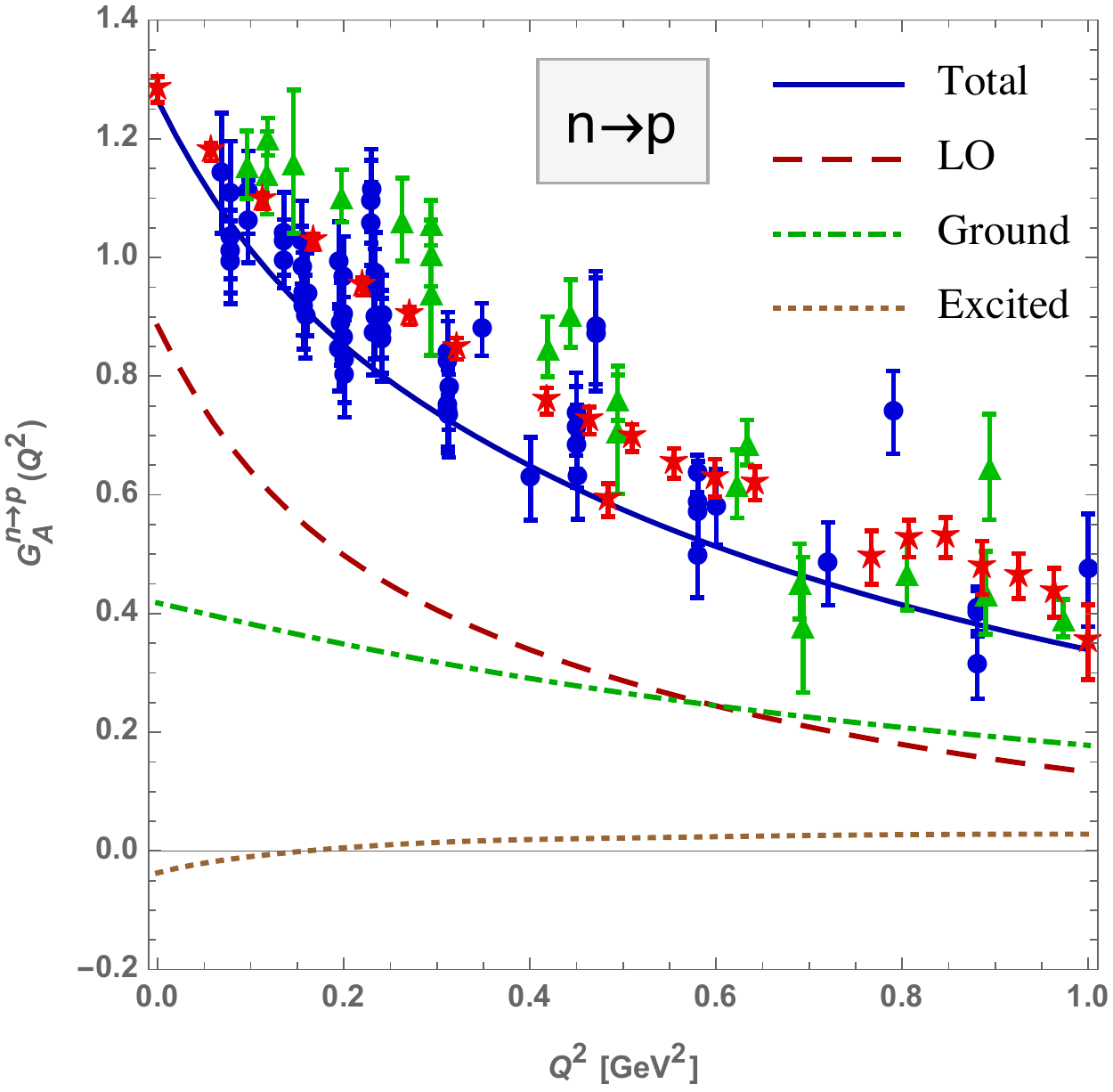}
\end{center}
\caption{\label{fig:Gnp} Result of the $n\to p$ transition axial form factor $G_A^{n\to p}(Q^2)$ comparing with pion electroproduction data (blue solid circles) and neutrino scattering data (green solid triangles)~\cite{Amaro:2016}, and the lattice QCD values (red stars) are taken from Ref.~\cite{Alexandrou:2021}.}
\end{figure}

In Fig.~\ref{fig:Gnp}, we present the $Q^2$ dependence of the $n\to p$ transition axial form factor $G_A^{n\to p}(Q^2)$ in the low energy region $Q^2\leqslant 1$ $\rm GeV^2$, and include the separate contribution from the three-quark core (dashed line), the meson cloud with the ground-state quark propagator (dot-dashed line) as well as the meson cloud with the excited states in the quark propagators (dotted line). For comparison, the pion electroproduction (blue solid circles) and the neutrino scattering (green solid triangles) experimental data~\cite{Amaro:2016} are plotted. We also present the recent lattice QCD simulation values (red stars), in which the simulations have been reproduced directly with the physical pion masses~\cite{Alexandrou:2021}. It is clear that the total result for the $G_A^{n\to p}(Q^2)$ in Fig.~\ref{fig:Gnp} is consistent with the pion electroproduction experimental data, but it exhibits a slight steep $Q^2$ dependence compared with the neutrino scattering data and the lattice QCD values.

As the PCQM results show in Fig.~\ref{fig:Gnp}, the LO gives a greater contribution to the $G_A^{n\to p}(Q^2)$ and results in a dipolelike form, and the meson cloud also contributes vastly to the $G_A^{n\to p}(Q^2)$ and mainly through the loop diagrams with the ground-state quark propagator, but the excited states in the intermediate quark propagators contribute rather less. One may also find that the meson cloud leads to a flat contribution to the $G_A^{n\to p}(Q^2)$. The flat contribution indicates that the sea quarks distribute mainly in a very small region, which is in accordance with the  finding of Ref.~\cite{Hammer:2004}.

In terms of the dipole fit for the $G_A$, the axial mass could be expressed~\cite{Bernard:2002} by 
\begin{equation}
\frac{12}{M_A^2}=-6\frac{1}{G_A(0)} \frac{\rm{d} G_A(Q^2)}{\rm{d} Q^2}|_{Q^2=0},\label{eq:MA}
\end{equation}
thus we may determine the nucleon axial mass $M_A$ in the PCQM as $M_A=1.157$ GeV, which is close to one adjusted from the pion electroproduction data, $1.069\pm 0.016$ GeV. Also, our result on the $M_A$ is in agreement with the one extracted from the MINER$\nu$A data by the spectral function approach, $1.13\pm0.06 \leqslant M_A \leqslant 1.16\pm 0.06$ GeV~\cite{Ankowski:2015}.

\begin{figure*}[t]
\begin{center}
\includegraphics[width=0.33\textwidth]{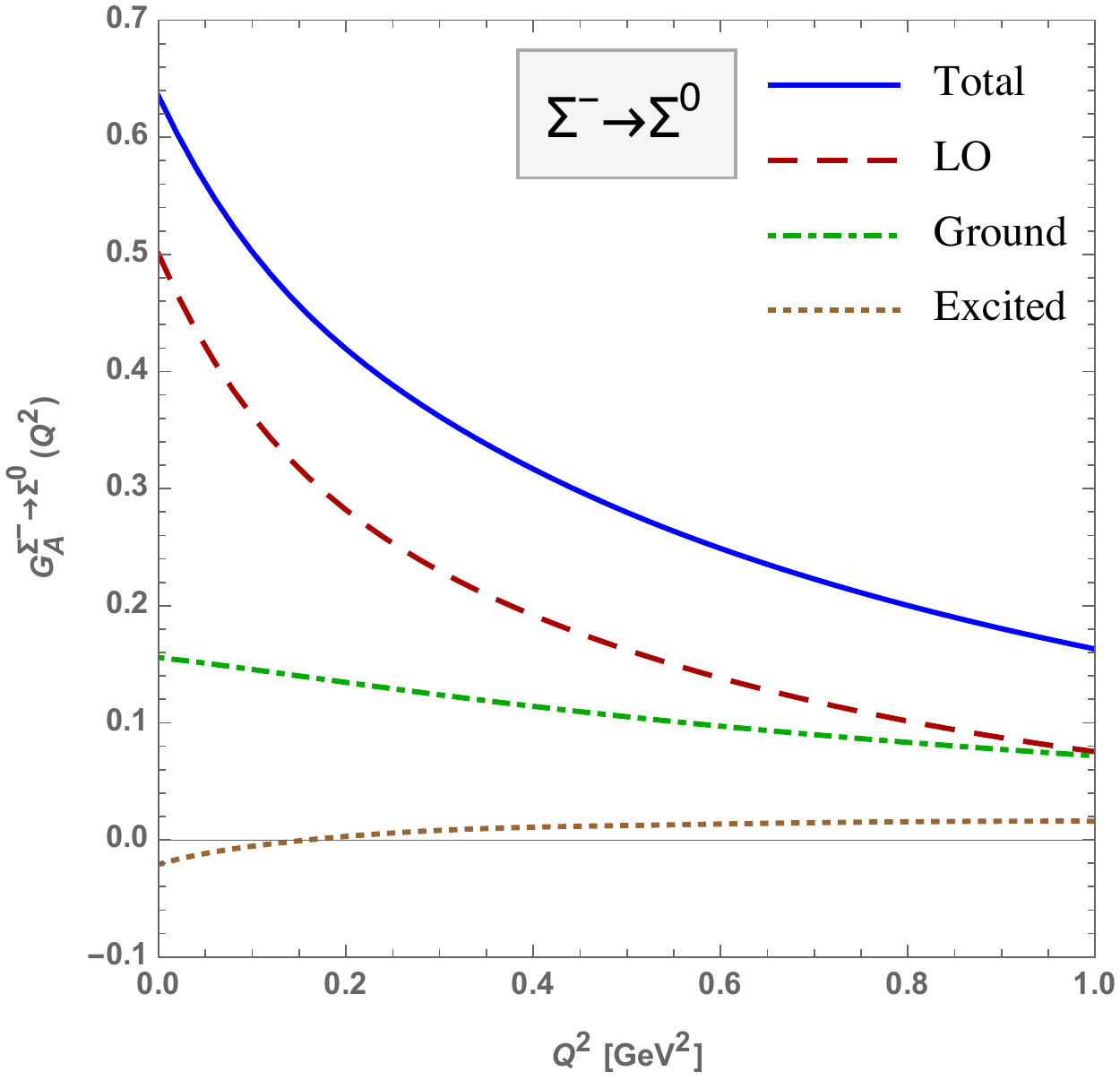}
\includegraphics[width=0.33\textwidth]{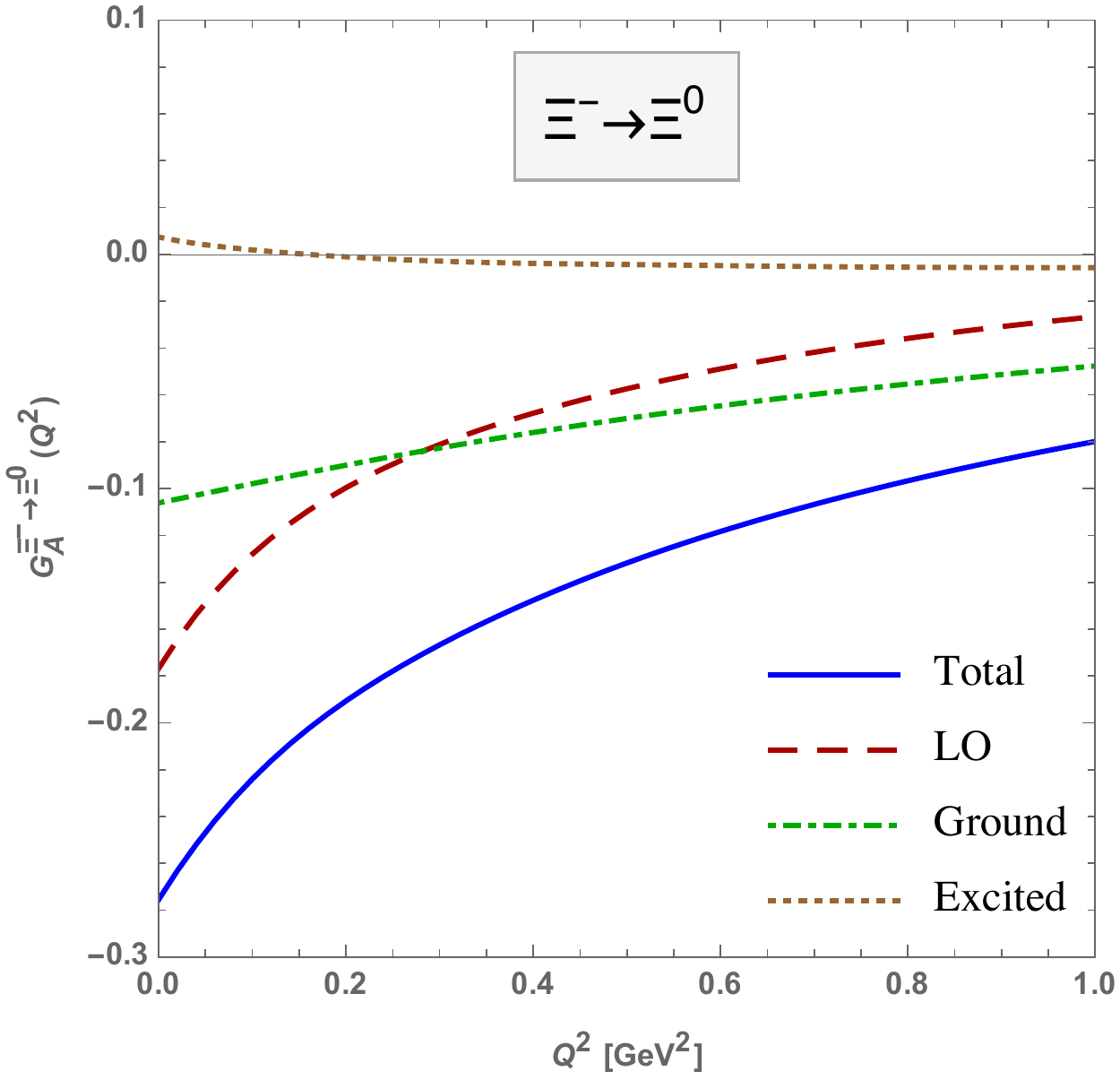}
\includegraphics[width=0.329\textwidth]{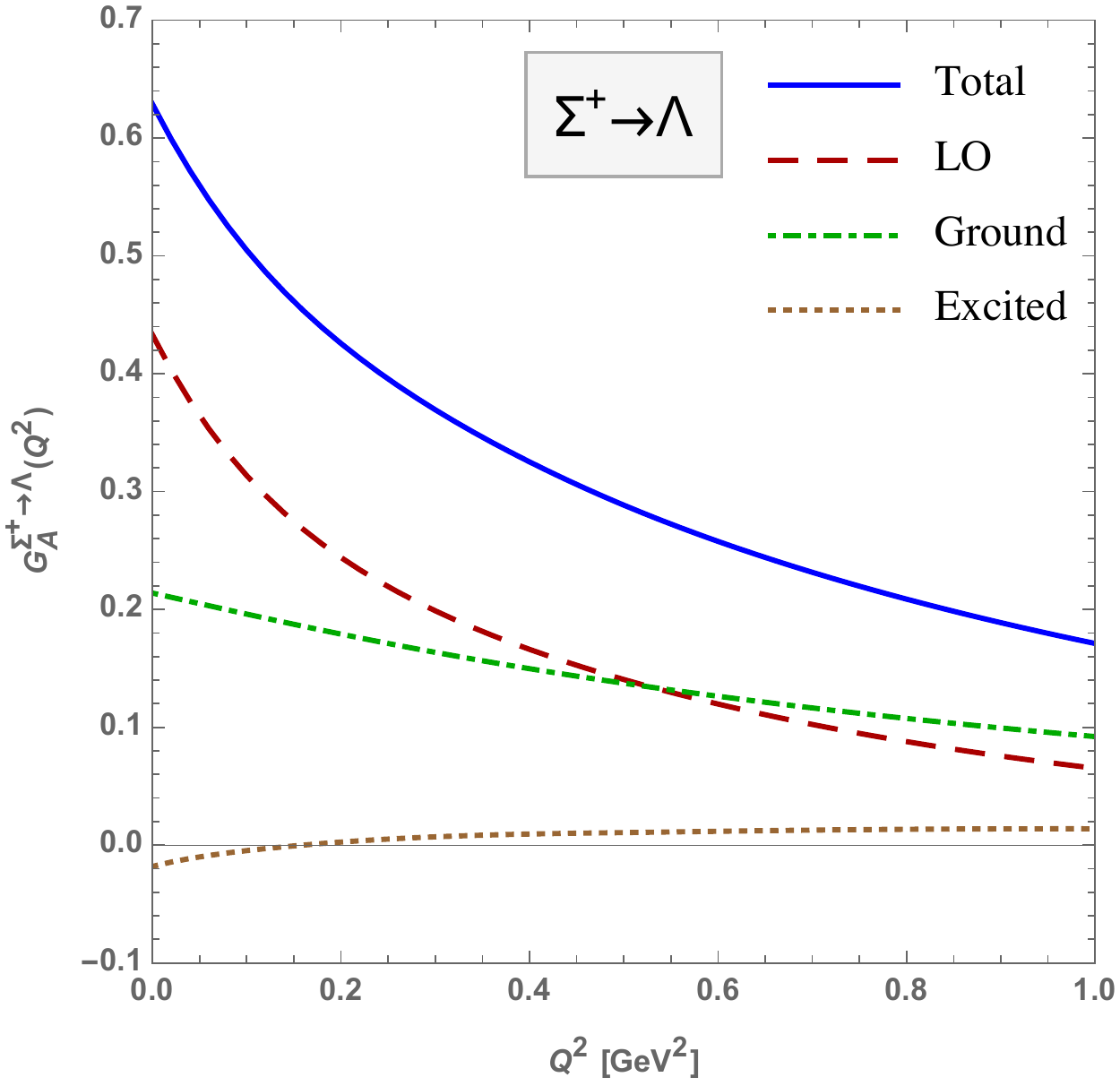}
\includegraphics[width=0.325\textwidth]{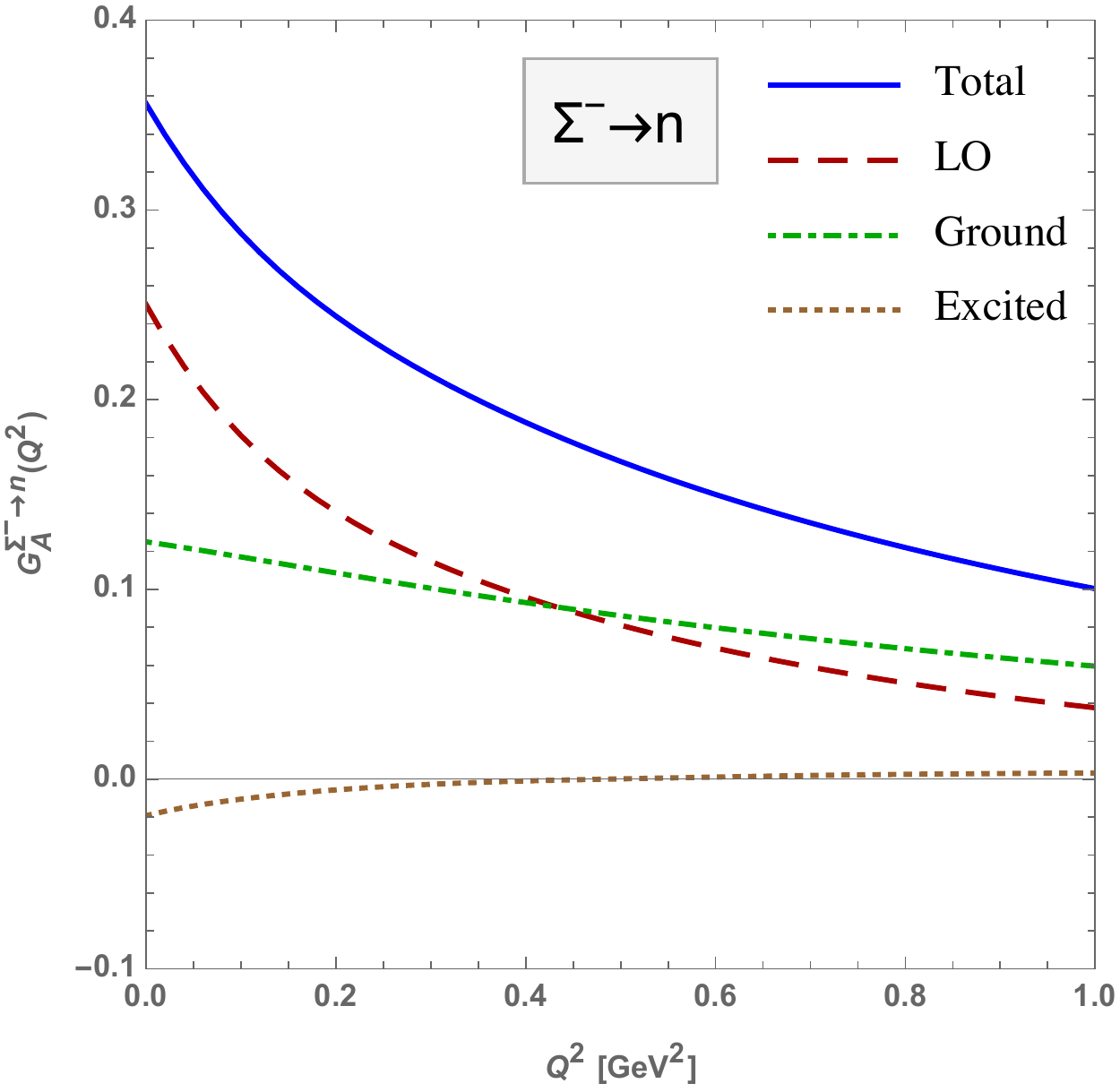}
\includegraphics[width=0.336\textwidth]{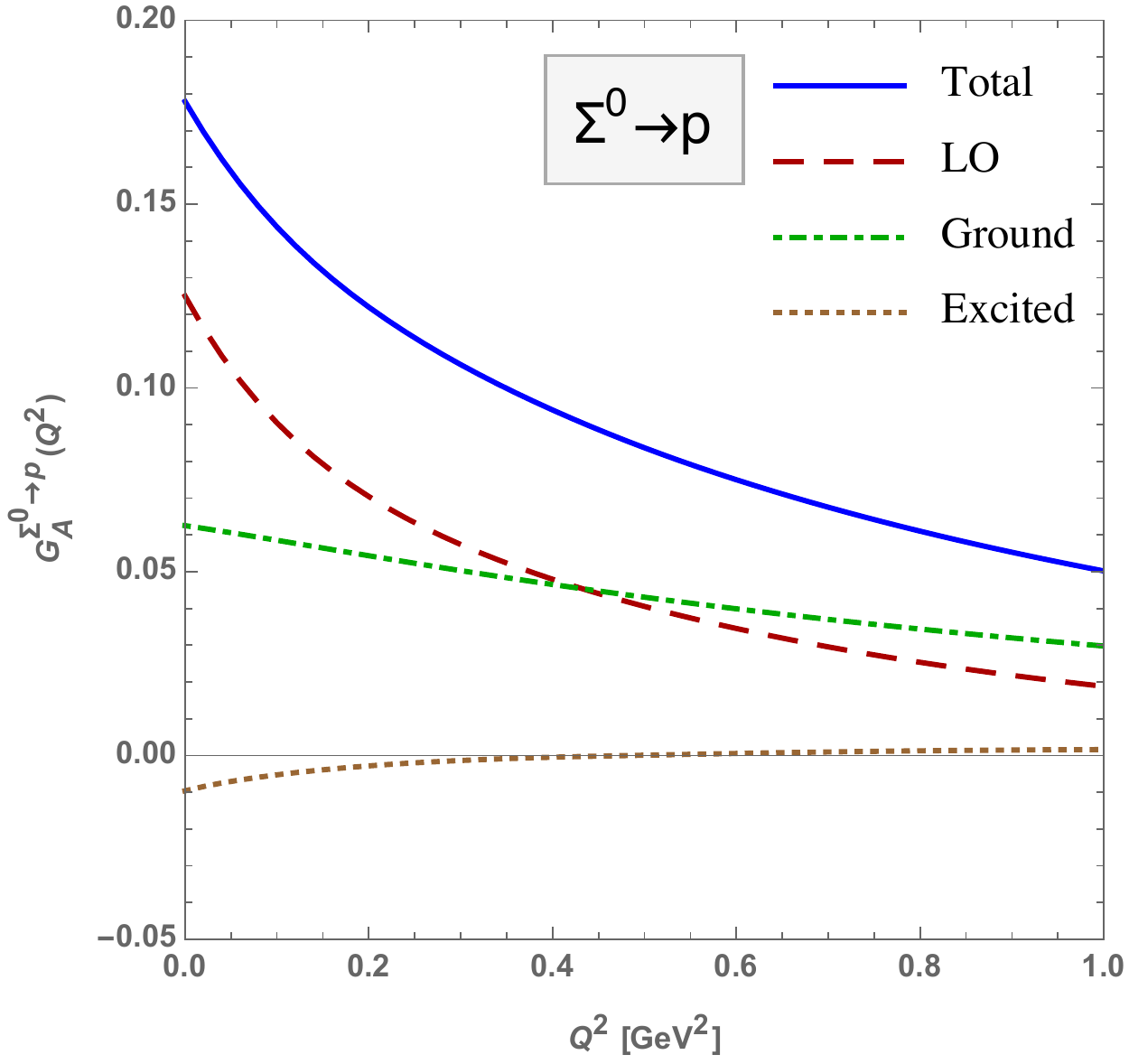}
\includegraphics[width=0.328\textwidth]{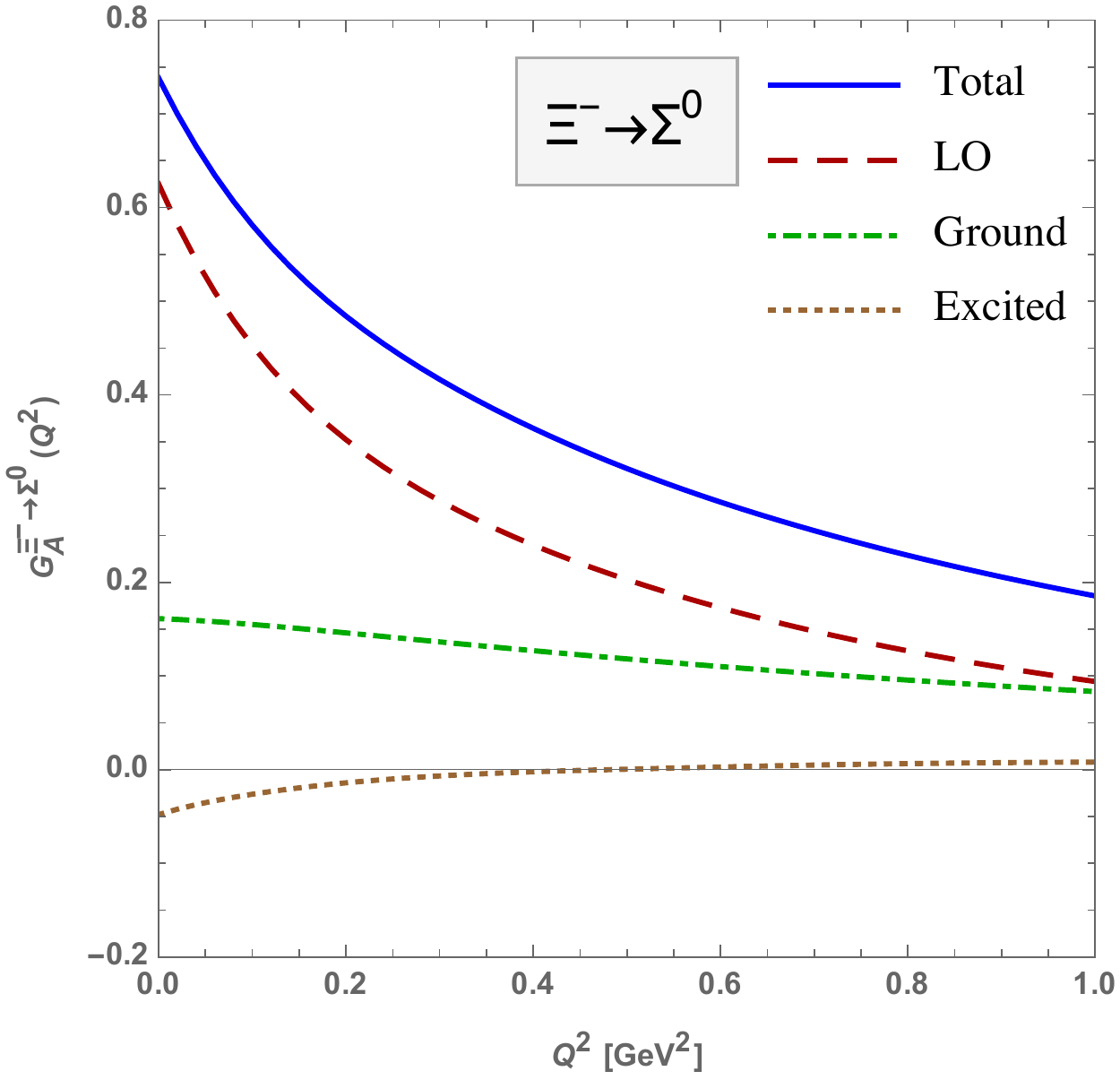}
\includegraphics[width=0.331\textwidth]{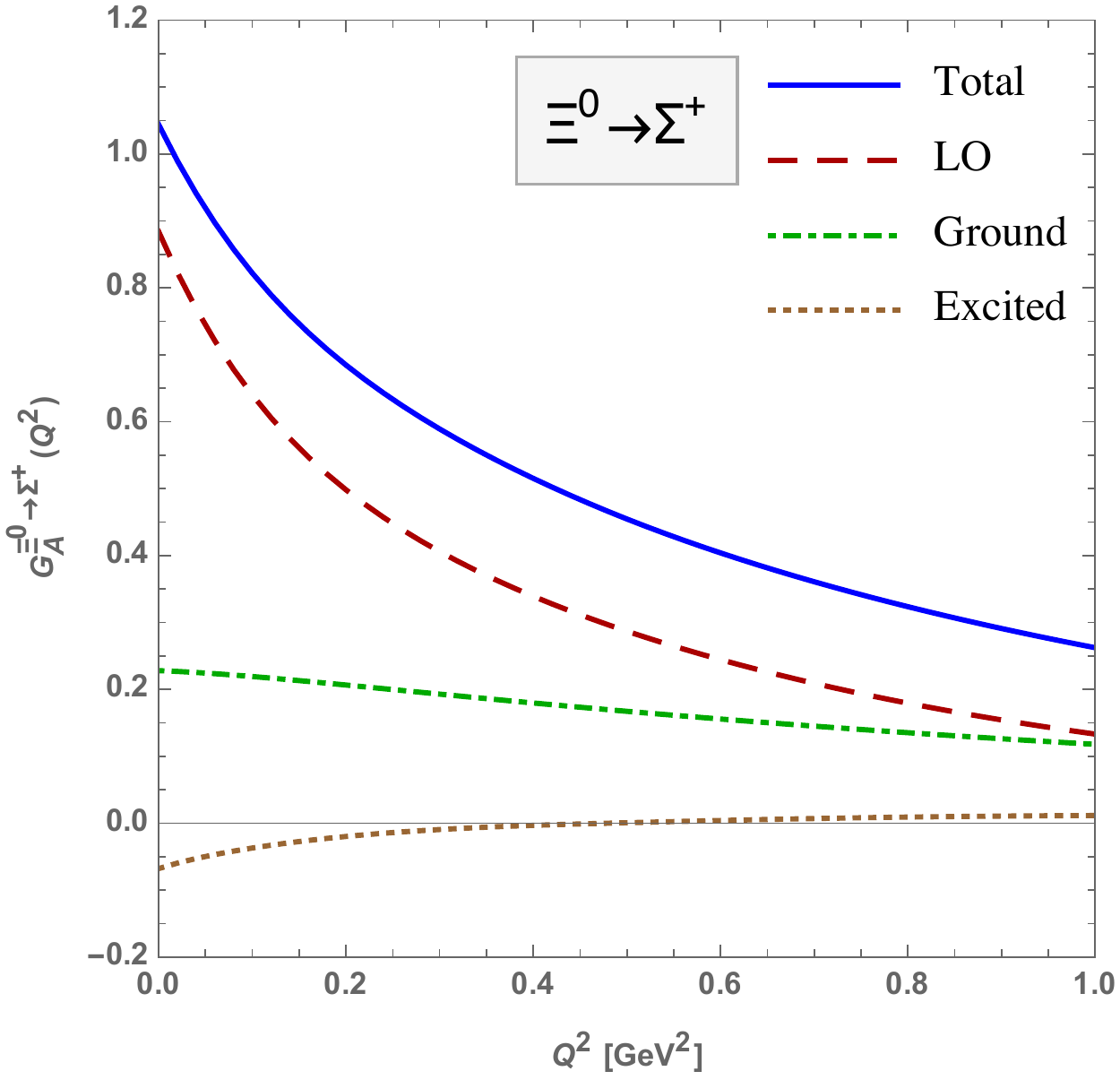}
\includegraphics[width=0.327\textwidth]{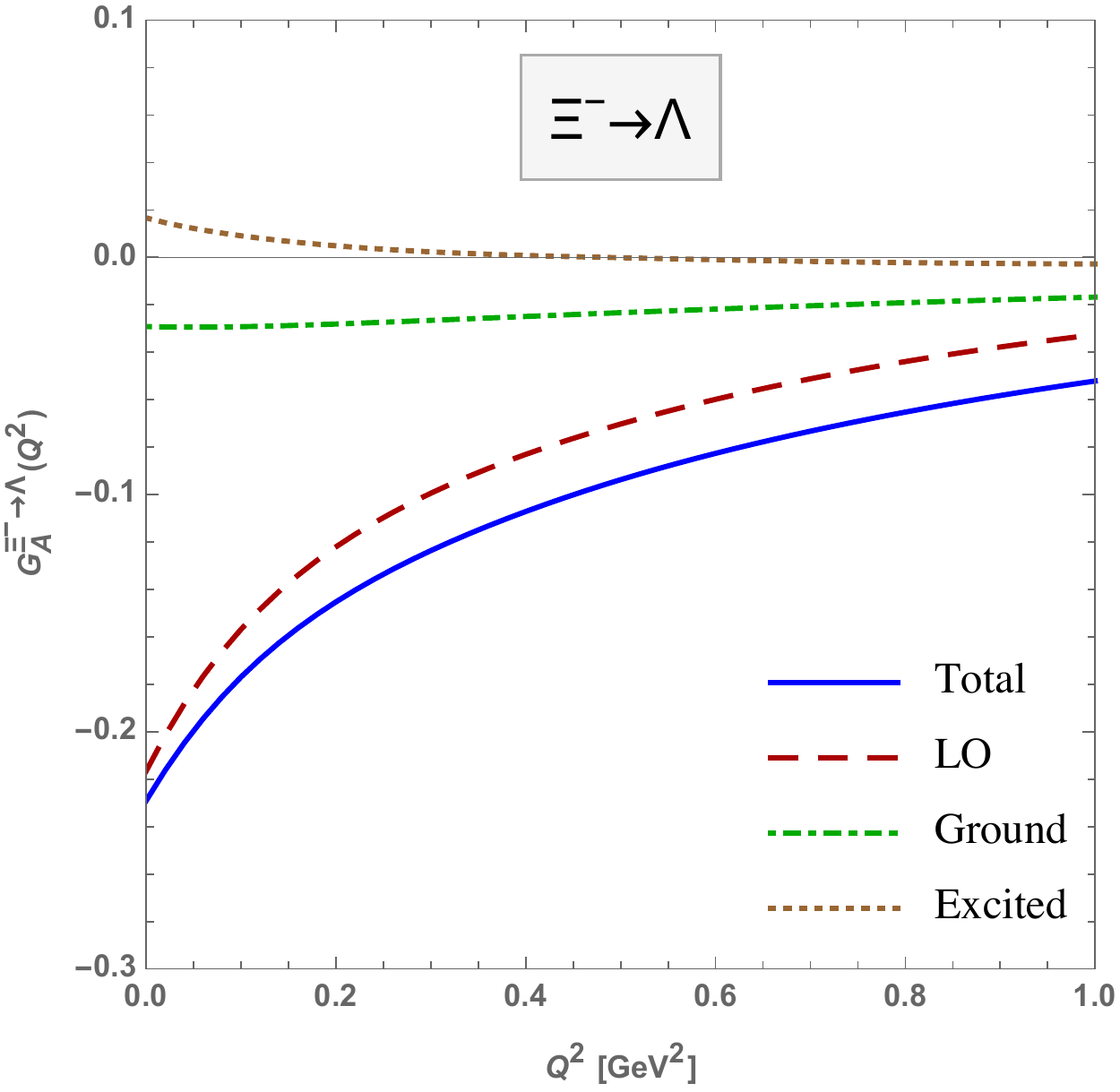}
\includegraphics[width=0.33\textwidth]{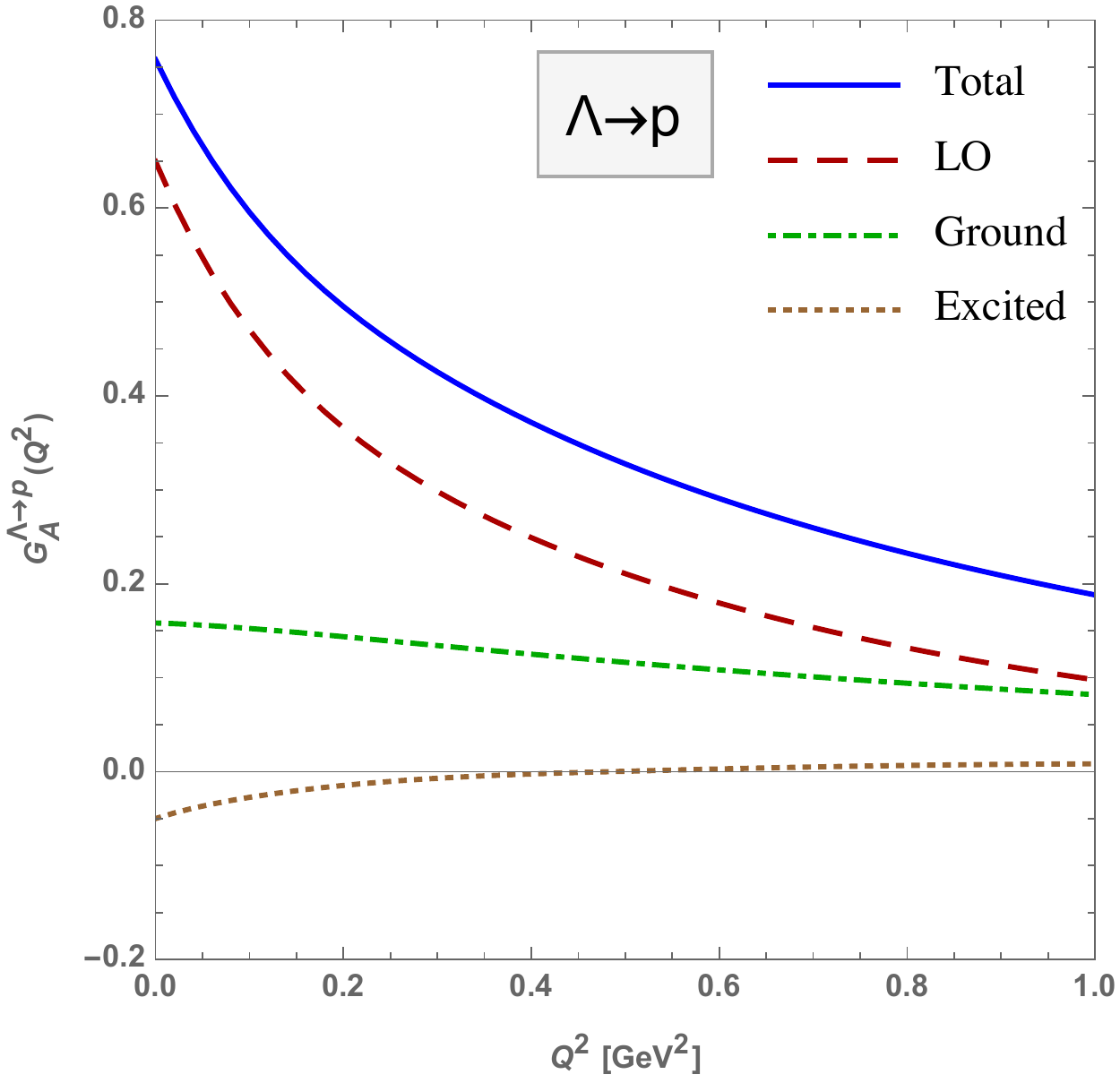}
\end{center}
\caption{\label{fig:GAB} Results of the $G_A^{B\to B'}(Q^2)$ for the various transitions}
\end{figure*}

Finally, we exhibit in Fig.~\ref{fig:GAB} a series of the PCQM predictions on the $G_A^{B\to B'}(Q^2)$ associated with the various other transitions, and again, the contributions from the 3q-core (dashed line), from the ground-state quark propagator (dot-dashed line) and from the excited-state quark propagators (dotted line) are included. It is seen in Fig.~\ref{fig:GAB} that the PCQM prediction on each $G_A^{B\to B'}(Q^2)$ shows a similar $Q^2$-dependence behavior with the one of the $G_A^{n\to p}(Q^2)$. Thus we may summarize that the $G_A^{B\to B'}(Q^2)$, including $n\to p$ transition, distributes as a dipolelike pattern, and it is caused by the 3q-core mainly. The meson cloud with the ground-state quark propagator is also very important and turns out a flat contribution, but the contribution from the excited-state quark propagator is very limited. 

\section{\label{sec:Summary}Summary and conclusions}

In this work, we have investigated and predicted the axial transition form factor $G_A^{B\to B'}(Q^2)$ and the axial transition charges $g_A^{B\to B'}$ in the framework of the PCQM for the various octet baryon transitions, such as $n\to p$, $\Sigma^- \to \Sigma^0$, $\Xi^- \to \Xi^0$, $\Sigma^+ \to \Lambda$, $\Sigma^-\to n$, $\Sigma^0 \to p$, $\Xi^- \to \Sigma^0$, $\Xi^0 \to \Sigma^+$, $\Xi^- \to \Lambda$ and $\Lambda\to p$. The calculations are extended to the SU(3) flavor symmetry, and include both the ground and excited states in the intermediate quark propagators. The ground and excited quark wave functions have been derived in our previous works by solving the Dirac equation with the Cornell-like potential numerically. Therefore, there is not any free parameter in the present work. 

In summary, one may conclude that the PCQM results on the $G_A^{B\to B'}(Q^2)$ and the $g_A^{B\to B'}$ agree well with the existing experimental data and the lattice-QCD estimates. The study shows clearly that the $G_A^{B\to B'}(Q^2)$ for all the $\Delta I=1$ and $\Delta S=1$ transitions behave a dipolelike form and stem mainly from the 3q-core. The meson cloud with the ground-state quark propagator plays an extremely important role in the $G_A^{B\to B'}(Q^2)$, while the excited-state quark propagator contributing to the $G_A^{B\to B'}(Q^2)$ could be regarded as the higher order corrections and it is very limited. The study also reveals that the sea quarks distribute in a very small region, as indicated by the flat contribution from the meson cloud.

\begin{acknowledgments}
This research has received funding support from the National Science Research and Innovation Fund (NSRF) via the Program Management Unit for Human Resources \& Institutional Development, Research and Innovation (grant number B05F640055). X. Y. L. and A. L. acknowledge support from Suranaree University of Technology (SUT) and the Office of the Higher Education Commission (CHE) under the National Research University (NRU) project of Thailand (SUT-CHE-NRU Contract No. FtR 09/2561). This work is also supported by the Scientific Research Foundation of Liaoning Province of China.
\end{acknowledgments}

\bibliography{Refs}

\end{document}